\documentclass[prd,eqsecnum,twocolumn,amsfonts,amssymb,showpacs]{revtex4}

\usepackage{CJK}

\usepackage{graphicx}

\usepackage{bm}

\newcommand{\beq}{\begin{equation}}
\newcommand{\eeq}{\end{equation}}
\newcommand{\beqs}{\begin{eqnarray}}
\newcommand{\eeqs}{\end{eqnarray}}

\newcommand{\pslash}{p\hspace{-0.067in}\slash}

\begin{document}

\begin{CJK*}{UTF8}{}

\title{Revisiting radiative decays of $1^{+-}$ heavy quarkonia in the
covariant light-front approach}

\author{Yan-Liang Shi (\CJKfamily{bsmi}石炎亮)}

\affiliation{C. N. Yang Institute for Theoretical Physics,
Stony Brook University, Stony Brook, N. Y. 11794 }

\begin{abstract}

  We revisit the calculation of the width for the radiative decay of a $1^{+-}$
  heavy $Q \bar Q$ meson via the channel $1^{+-} \to 0^{-+} +\gamma$ in the
  covariant light-front quark model.  We carry out the reduction of the
  light-front amplitude in the non-relativistic limit, explicitly computing the
  leading and next-to-leading order relativistic corrections.  This shows the
  consistency of the light-front approach with the non-relativistic formula for
  this electric dipole transition. Furthermore, the theoretical uncertainty in
  the predicted width is studied as a function of the inputs for the heavy
  quark mass and wavefunction structure parameter.  We analyze the specific
  decays $h_{c}(1P) \to \eta_{c}(1S) + \gamma$ and $h_{b}(1P) \to \eta_{b}(1S)
  + \gamma$.  We compare our results with  experimental data and with other
  theoretical predictions from calculations based on non-relativistic models
  and their extensions to include relativistic effects, finding reasonable
  agreement.

\end{abstract}

\pacs{13.20.-v, 13.20.Gd, 12.39.Ki}

\maketitle

\end{CJK*}




\section{Introduction}

Heavy-quark $Q \bar Q$ bound states play a valuable role in elucidating the
properties of quantum chromodynamics (QCD).  Since the discoveries of the
$J/\psi$ in 1974 \cite{Aubert:1974js,psi_slac} and other $c \bar c$ charmonium states, and the $\Upsilon$ in
1977 \cite{Herb:1977ek,Innes:1977ae} and other $b \bar b$ states, we now have a very substantial set of data on
the properties and decays of these quarkonium states.  Some reviews include
\cite{quiggrosner79}-\cite{rosner2013}.  The goal of understanding these data
motivates theoretical studies, in particular, studies of the decays of $Q \bar
Q$ states.

Among various decay channels, radiative decays are a very
good testing ground for models, since the emitted photon is directly detected
and the electromagnetic interaction is well understood. An electric dipole (E1)
transition is one the simplest types of radiative decays. Here we consider E1
transitions of the form
\beq
{}^{1}P_{1} \to {}^{1} S_{0} + \gamma ,
\label{heta}
\eeq
where a spin-singlet P-wave $Q \bar Q$ quarkonium state decays to a
spin-singlet S-wave $Q \bar Q$ state.  In terms of the spin $J$ and the charge
and parity quantum numbers P and C, indicated as $J^{PC}$, this has the form
$1^{+-} \to 0^{-+} + \gamma$.

Several theoretical analyses of these E1 transition rates have been carried
out, using various models \cite{Karl:1980wm}-\cite{Li:2009zu}.  A number of
these models utilize the non-relativistic quantum mechanics formula for an E1
transition, involving the calculation of the overlap integral of the quarkonium
wavefunctions of the initial and final states. The quarkonium wavefunction is
obtained from the solution of the Schr\"odinger equation with non-relativistic
potentials, such as the Cornell potential, $V = -(4/3)\alpha_s(m_Q)/r + \sigma
r$. The first term in this potential is a non-Abelian Coulomb potential
representing one-gluon exchange at short distances, where
$\alpha_s(m_Q) = g_s(\mu)^2/(4\pi)$ is the strong coupling evaluated at the
scale of the heavy quark mass, $m_Q$, and the second term is the
linear confining potential, where $\sigma = 0.18$ GeV$^2$ is the string
tension.  Current data yield a fit to
$\alpha_s(\mu)$ such that $\alpha_s \simeq 0.33$ at the scale $\mu = 1.5$ GeV
relevant for $c \bar c$ states and $\alpha_s \simeq 0.21$ at the scale $\mu =
4.7$ GeV relevant for $b \bar b$ states \cite{PDG}.  Relativistic corrections
have also been calculated by replacing the Schr\"odinger equation by the Dirac
equation, and computing corrections in powers of $v/c$, where $v$ is the
velocity of the heavy (anti)quark in the rest frame of the $Q \bar Q$ bound
state.

It is of interest to study the radiative decays (\ref{heta}) with a fully
relativistic approach, namely the light-front quark model
(LFQM)\cite{Terentev:1976jk}-\cite{Cheng:2003sm}. This approach naturally
includes relativistic effects of quark spins and the internal motion of the
constituent quarks. Another advantage of the light-front quark model is that it
is manifestly covariant.  Hence it is easy to boost a hadron bound state from
one inertial Lorentz frame to another one when the bound state wavefunction is
known in a particular frame\cite{Brodsky:1997de}.  The light-front approach has
been used to study semileptonic and nonleptonic decays of heavy-flavor $D$ and
$B$ mesons and also to evaluate radiative decay rates of heavy mesons
\cite{Hwang:2006cua,Choi:2007se,Hwang:2010iq,Ke:2010vn,Ke:2013zs,Shi:2016nxr}.

In this paper, we extend our previous work with Ke and Li in
Ref.\cite{Ke:2013zs} on the study of the radiative decays
\beq
h_{c}(1P) \to \eta_{c}(1S) + \gamma \ ,
\eeq
and
\beq
h_{b}(1P) \to \eta_{b}(1S) + \gamma \ .
\eeq
We present several new results here.  We carry out the reduction of the
light-front amplitude to the non-relativistic limit, explicitly computing the
leading and next-to-leading order relativistic corrections.  This shows the
consistency of the light-front approach with the non-relativistic formula for
this electric dipole transition. Furthermore, we investigate the theoretical
uncertainties in the predicted widths as functions of the inputs
for the heavy quark mass and wavefunction structure parameters.
As in Ref. \cite{Ke:2013zs}, we compare our numerical results for these widths
with experimental data and with other theoretical predictions from
calculations based on non-relativistic models and their extensions to include
relativistic effects, extending \cite{Ke:2013zs} with further study of the
theoretical uncertainties in our calculations. Specifically, we
compare our numerical results with results from
\cite{Voloshin:2007dx,Brambilla:2004wf,Ebert:2002pp,Li:2009nr,Godfrey:2015dia,Segovia:2016xqb,Barnes:2005pb,Li:2009zu}
as well as latest experimental data \cite{PDG}.

The paper is organized as follows: In Section \ref{LFQM}, we review the
formulas for the radiative decay $1^{+-} \to 1^{-+} +\gamma$. In section
\ref{NR}, we analyze the reduction of light-front formulas when applied to
heavy quarkonium systems to non-relativistic limit and compare these with
non-relativistic quantum mechanical electric dipole transition formula.  In
Section \ref{ANALYSIS}, we present our numerical results for the decay widths
of $h_{c}(1P) \to \eta_{c}(1S) + \gamma $ and $h_{b}(1P) \to \eta_{b}(1S) +
\gamma $ including an extended analysis of the theoretical uncertainties.
Our conclusions are given in Section \ref{CON}.


\section{Light-front formalism for the decays $1^{+-}  \to 0^{-+} +\gamma$}
\label{LFQM}

\subsection{Notation}

We first define some notation, retaining the conventions of
\cite{Jaus:1999zv,Cheng:2003sm}. In light-front
coordinates, the four-momentum $p$ is
\beq
p^\mu = (p^{-},p^{+},{\bf p}_{\perp})
\eeq
where $p^\pm = p^0 \pm p^3$ and ${\bf p}_{\perp}=(p^1,p^2)$.
Hence, the Lorentz scalar product $p^2 = p_\mu p^\mu$ is
\beqs
p^2 & = & (p^0)^2 - |{\bf p}|^2 = (p^0)^2 - (p^3)^2 - |{\bf p}_\perp|^2 \cr\cr
    & = & p^+ p^- -|{\bf p}_\perp|^2 \ .
\eeqs

Consider a decay of a $Q \bar Q$ meson consisting of two constituent particles
(quark and antiquark).  The momentum of the parent meson is denoted as $P'=p'_1
+p_2$, where $p'_{1}$ and $p_2$ are the momenta of the constituent quark and
antiquark, with mass $m'_1$ and $m_2$, respectively. The momentum of the
daughter $Q \bar Q$ meson is written as $P''=p''_1+p_2$, where $p''_{1}$ is the
momentum of the constituent quark, with mass $m''_1$. Here we have $m'_1=m_2 =
m''=m_Q$.  The four-momentum of the parent meson with mass $ M'$ can be
expressed as $P'=(P'^{-},P'^{+},{\bf P}'_{\perp})$, where
$P'^2=P'^{+}P'^{-}-|{\bf P'}_{\perp}|^2=M'^2$. Similarly, for the daughter
meson with mass $M''$, one has $P''^2=M''^2$, as shown in Fig. \ref{p1} below.
(Vector signs on transverse momentum components are henceforth taken to be
implicit.)

The momenta of the constituent quark and antiquark ($p'_1$, $p''_1$ and $p_2$)
can be described by internal variables $(x_2,p'_{\perp})$ thus:
\beqs
&&p'^{+}_1=x_1P'^{+}, \quad  p^{+}_2=x_2P'^{+} \nonumber\\
&&p'_{1\perp}=x_1P'_{\perp}+p'_{\perp}, \quad
  p_{2\perp}=x_2P'_{\perp} - p'_{\perp} \nonumber\\
&&x_1 + x_2 =1 \ .
\eeqs
Explicitly,
\beq
x_1= \frac{e_1- p'_z }{e'_1+e_2} \ , \quad  x_2= \frac{p'_z+e_2}{e'_1+e_2} \ ,
\eeq
where $e'_1$,  $e''_1$ and $e_2$ are the energy of the quark (anti-quark)
with momenta $p'_1$,  $p''_1$ and  $p_2$:
\beqs
e'_1 &=&\sqrt{m'^2_1+p'^2_{\perp}+p'^2_{z}} \nonumber \\
e''_1 &=&\sqrt{m''^2_1+p''^2_{\perp}+p''^2_{z}} \nonumber \\
e_2 &=&\sqrt{m^2_2+p'^2_{\perp}+p'^2_{z}} \ .
\eeqs
With the external momentum of the photon given as $q=P'-P''$,
$p''_{\perp}$ can be expressed as
\beq
p''_{\perp}=p'_{\perp}-x_2 q_{\perp} .
\eeq
Here $p'_z$ and $p''_z$ can also be expressed as functions of internal
variables $(x_2,p'_{\perp})$, and explicit expressions can be found in Appendix
\ref{expression}.


\subsection{Form factors}

Define external momentum variables to be $P=P'+P''$, $q=P'-P''$, where $q$ is
the four-momentum of the photon that is emitted in the radiative transition.
The general amplitude of the radiative decay (\ref{heta}) of the axial vector
$1^{+-}$ ${}^1P_{1}$ meson, denoted as $A$, to the pseudoscalar
$0^{-+}$ ${}^1S_{0}$ meson, denoted as $P$, can be
written as \cite{Cheng:2003sm}:
\beq i{\cal A} \left( A (P') \to P(P'')
  \gamma (q) \right) = \varepsilon^{*}_{\mu}(q) \varepsilon'_{\nu}(P')i \tilde
{\cal A }^{\mu\nu} \ ,
\label{generalamp}
\eeq
where
\beq
i \tilde {\cal A }^{\mu\nu} = f_1(q^2) g^{\mu\nu}+  P^{\mu}\left [f_{+}(q^2) P^{\nu}  +  f_{-} (q^2) q^{\nu} \right ] \ .
\eeq
In the above expression, we have used the condition
$\varepsilon^{*}_{\mu}(q)q^{\mu}=0 $ to eliminate terms that are proportional
to $q^\mu$. This expression can be simplified further by using the
transversality property of axial vector polarization vector:
\beq
\varepsilon'_{\nu}(P')(P+q)^{\nu}=0 \ .
\eeq
Then the general amplitude can be written as
\beq
i \tilde {\cal A }^{\mu\nu} \to i  {\cal A }^{\mu\nu} = f_1(q^2) g^{\mu\nu}+  f_2(q^2) P^{\mu} q^{\nu} \ ,
\label{amplit}
\eeq
where $f_2(q^2)$ is linear combination of $f_{+}(q^2)$ and  $f_{-} (q^2)$:
\beq
f_2(q^2)= - f_{+}(q^2) +  f_{-} (q^2) \ .
\eeq
Notice that in Eq.(\ref{amplit}), $f_1(q^2)$ and $ f_2(q^2)$ are not
independent. Because of electromagnetic gauge invariance, they are related by
the following equation:
\beq
q_{\mu}  {\cal A }^{\mu\nu}=0  \ \to \  f_1(q^2)+ f_2(q^2) (P \cdot q ) =0 \ .
\eeq
We find that the $f_2(q^2)$ term in the amplitude (Eq.(\ref{amplit})) does not
contribute to the width, due to the fact that the photon has only two
transverse polarization states, so that the timelike component
$\varepsilon^{*}_{0}(q)$ is equal to zero. Taking the parent axial vector meson
$A(P')$ in its rest frame, we have $P'_{\nu}=(M',0)$, where $M'$ is the mass of
the vector meson $A(P')$. The contribution of the $f_2 (q^2)$ term is zero
after we contract ${\cal A }^{\mu\nu} $ with polarization vector
$\varepsilon^{*}_{\mu}(q)$:
\beq
\varepsilon^{*}_{\mu}(q) \varepsilon'_{\nu}(P') [f_2(q^2) P^{\mu} q^{\nu}] \propto \varepsilon^{*}_{\mu}(q) P'^{\mu}= \varepsilon^{*}_{0}(q) P'^{0} =0 \ .
\eeq
That fact that $f_2(q^2)$ has no contribution can also be shown
straightforwardly when we calculate the transition probability $|{\cal A}|^2$,
using exact summation of two physical polarizations of photon:
\beq \sum
\varepsilon^{*}_{\mu}(q) \varepsilon^{}_{\alpha}(q)= - g_{\mu\alpha} +
\frac{q_{\mu}\eta_{\alpha}+ q_{\alpha}\eta_{\mu} }{2 q\cdot \eta} \ ,
\eeq
where $q= ( |{\bf q}| , {\bf q} )$ and $\eta = ( |{\bf q}| , - {\bf q} )$. After
an explicit calculation, we have
\beq \sum_{\text{ polarization}}|{\cal A}|^2 = 2|f_1 (q^2)|^2 \ .
\eeq
Taking the physical value $q^2 \to 0$ in the form factor $f_1 (q^2)$ and
averaging initial state polarizations, the radiative transition width of
$1^{+-} \to 0^{-+}+ \gamma$ is given by
\beq \Gamma=\frac{1}{3}\cdot
\frac{|{\bf q}|}{8\pi M'^2} \sum_{\text{polar.}} |{\cal A}|^2 = \frac{|\bf
  {q}|}{12 \pi M'^2} \cdot |f_1 (0)|^2 \ ,
\label{width}
\eeq
where the energy of the emitted photon is related to the masses of mesons
as $|{\bf q}|= (M'^2-M''^2)/(2M')$.

\begin{figure}
\begin{center}
\resizebox{0.45\textwidth}{!}{%
\includegraphics{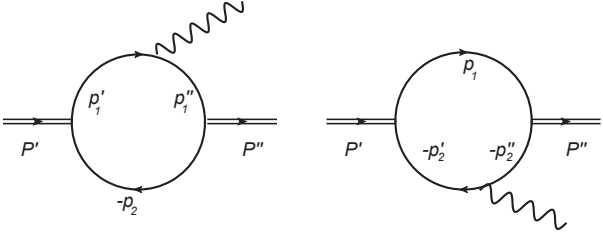}
}%
\caption {Feynman diagrams for radiative transition $1^{+-} \to 0^{-+}+\gamma$
  in the light-front approach.}
\label{p1}
\end{center}
\end{figure}


\subsection{Calculation of radiative decay amplitude}

In the covariant light-front quark model, the vertex function of the axial
vector meson $A$ ($1^{+-}$,${}^1P_{1}$) is given by
\beq
-iH'_A \left[\frac{1}{W'_{A}}(p'_1-p_2)^{\mu}\right]\gamma^5 \ ,
\eeq
and the vertex function of the pseudoscalar meson $P$ ($0^{-+}$,${}^1S_{0}$) is
given by
\beq
H''_P \gamma^5 \ ,
\eeq
where $H'_A$ and $H''_P$ are functions of $p'_1$ and $p_2$, and $W'_{A}$ can be
reduced to a constant, which we will discuss later in this subsection.

In the light-front framework that we use \cite{Jaus:1999zv,Cheng:2003sm}, at
leading order there are two diagrams that contribute to the $A \to P + \gamma $
transition amplitude.  These give the corresponding contributions to this
amplitude
\beq i {\cal A}^{\mu\nu} (A \to P
+ \gamma ) = i {\cal A}^{\mu\nu} (a) + i {\cal A}^{\mu\nu} (b)
\eeq
where
$i{\cal A}^{\mu\nu} (a) $ and $i{\cal A}^{\mu\nu} (b) $ correspond to the left
and right diagram in Fig.(\ref{p1}), respectively. The contribution to the
amplitude from the right diagram can be obtained by taking the charge
conjugation of left diagram (see also \cite{Shi:2016nxr}). So we discuss the
left-hand diagram, where the corresponding transition amplitude is given by
\beq i {\cal A}^{\mu\nu} (a)= i \frac{e N_{e'_1}N_c}{(2\pi)^4} \int d^4p'_1
\, \frac{H'_A H''_P}{N'_1 N''_1 N_2} {\cal S}^{\mu\nu}_a \ , \eeq
where
\beqs
{\cal S}^{\mu\nu}_a &=& \text {Tr} [\gamma^5 (\pslash''_1+m''_1)\gamma^{\mu}(\pslash'_1+m'_1)\gamma^5(-\pslash_2+m_2)] \nonumber\\
  &\times&\frac{1}{W'_{A}}(2p'_1 -\frac{P+q}{2})^{\nu}  \nonumber\\
  &=& \frac{4}{W'_{A}}(2p'_1 -\frac{P+q}{2})^{\nu} [p''^{\mu}_1(p'_1\cdot p_2)+ p'^{\mu}_1(p''_1\cdot p_2)  \nonumber\\
  &-& p^{\mu}_2(p'_1\cdot p''_1) + m'_1 m_2 p''^{\mu}_1 + m''_1 m_2 p'^{\mu}_1 + m'_1 m''_1 p^{\mu}_2]  \nonumber\\
  &=& \frac{1}{W'_{A}}(2p'_1 -\frac{P+q}{2})^{\nu} \{ 2p'^{\mu}_1 [ M'^2+M''^2-q^2-2N_2  \nonumber\\
   &-& (m'_1 -m_2)^2-(m''_1-m_2)^2 +(m'_1-m''_1)^2] \nonumber\\
   &+& q^{\mu} [q^2-2M'^2+N'_1-N''_1+2N_2+2(m'_1-m_2)^2 \nonumber\\
   &-& (m'_1-m''_1)^2]+P^{\mu}[q^2-N'_1 -N''_1-(m'_1 - m''_1)^2] \}, \nonumber\\
   \label{diagamA}
\eeqs
\beqs
&&N'_1=p'^2_1-m'^2_1+i\epsilon, \nonumber\\
&&N''_1=p''^2_1-m''^2_1+i\epsilon, \nonumber\\
&&N_2=p^2_2-m^2_2+i\epsilon .
\eeqs
Here $N_{e'_1(e_2)}$ represents the electric charge of quark with four momentum
$p'_1$ ($p_2$). Here we have $N_{e'_1(e_2)}=e_{Q}$.
In Eq.(\ref{diagamA}), we have already applied the following relations:
\beqs
p''_1 &=& p'_1-q \nonumber\\
p_2 &=& (P+q)/2-p'_1 \nonumber\\
2p'_1\cdot p_2 &=& M'^2-N'_1-m'^2_1-N_2-m^2_2  \nonumber\\
2p''_1\cdot p_2 &=& M''^2-N''_1-m''^2_1-N_2-m^2_2   \nonumber\\
2p'_1\cdot p''_1 &=& -q^2 + N'_1+m'^2_1+N''_1+m''^2_1.
\label{relations}
\eeqs
Then we integrate over $p'^{-}_{1}$ by closing the contour in the upper complex
$p'^{-}_{1}$ plane, which amounts to the following the following replacement
\cite{Jaus:1999zv,Cheng:2003sm}:
\beqs
\int d^4p'_1 \,
\frac{H'_{A} H''_{P}}{N'_1N''_1 N_2} {\cal S}^{\mu\nu }_a \varepsilon^{*}_{\mu} \varepsilon'_{\nu}\to \nonumber \\
-i \pi \int dx_2 d^2p'_{\perp} \, \frac{h'_{A}h''_{P}}{x_2 \hat N'_1\hat
  N''_1}\hat{ \cal S}^{\mu\nu }_a \hat \varepsilon^{*}_{\mu} \hat
\varepsilon'_{\nu}  \ ,
\label{replacement}
\eeqs
where
\beqs
&&N'_1 \to \hat N'_1= x_1 (M'^2-M'^2_{0}) \nonumber \\
&&N''_1 \to \hat N''_1= x_1 (M''^2-M''^2_{0})  \nonumber \\
&& H'_{A} \to h'_{A}=(M'^2-M'^2_{0})  \sqrt{\frac{x_1 x_2}{N_c}} \frac{1}{\sqrt{2} \tilde M'_{0}} \varphi_{p} (p'_{\perp},x_2)\nonumber\\
&& H''_{P} \to h''_{P}=(M''^2-M''^2_{0})  \sqrt{\frac{x_1 x_2}{N_c}} \frac{1}{\sqrt{2}\tilde M''_{0}} \varphi (p''_{\perp},x_2) \nonumber\\
&&W'_{A} \to w'_{A} =  2 \ .
\label{replacementrelat}
\eeqs
In the above expressions, $\varphi_{p} (p'_{\perp},x_2)$ is the light-front
momentum space wavefunction for initial P-wave meson (${}^1P_{1}$), and
$\varphi (p''_{\perp},x_2)$ is the wavefunction for the final S-wave meson,
${}^1S_{0}$. Some details concerning the wavefunctions are given in Appendix
\ref{wavefunction}. The explicit forms of $M'_0$, $M''_0$, $\tilde M'_0$ and
$\tilde M''_0$ are listed in Appendix \ref{expression}. The definitions of
$\hat \varepsilon^{*}$, $\hat \varepsilon'$ and $\hat \varepsilon''^{*}_{\rho}$
are given in \cite{Jaus:1999zv,Cheng:2003sm}.

It is also necessary to include the contribution from zero modes in the $A$
meson.  This is equivalent to the following replacement for $p'_{1\mu}$
and $\hat N_2$ in $\hat{ \cal S}^{\mu\nu}_a $ in the integral
\cite{Jaus:1999zv,Cheng:2003sm}:
\beqs
\hat p'_{1\mu} &\to& P_{\mu} A^{(1)}_1 + q_\mu A^{(1)}_2 \ , \nonumber\\
 \hat p'_{1\mu} \hat p'_{1\nu}  &\to &g_{\mu\nu} A^{(2)}_1 + P_{\mu}P_{\nu} A^{(2)}_2   \ , \nonumber\\
&+& (P_{\mu}q_{\nu} + q_{\mu}P_{\nu})A^{(2)}_3 + q_{\mu} q_{\nu} A^{(2)}_4 \ , \nonumber \\
\hat p'_{1\mu} \hat N_2 &\to&  q_{\mu} [A^{(1)}_2 Z_2 + \frac{q\cdot P}{q^2}A^{(2)}_1]  \ , \nonumber\\
 \hat p'_{1\mu} \hat p'_{1\nu}  \hat N_2 &\to& g_{\mu\nu} A^{(2)}_1 Z_2 + q_{\mu}q_{\nu} [A^{(2)}_4 Z_2 + 2 \frac{q\cdot P}{q^2} A^{(1)}_2 A^{(2)}_1] , \nonumber\\
 \label{zeromode}
\eeqs
where the explicit expressions for $A^{(i)}_{j} (i,j=1 \sim 4) $ and $Z_2$ are
listed in Appendix \ref{expression}.

Combining Eq.(\ref{replacement}), Eq.(\ref{replacementrelat}) and
Eq.(\ref{zeromode}), we get
${\cal S}^{\mu\nu}_a \to \hat {\cal S}^{\mu\nu}_a $,
with the explicit form:
\beqs
\hat {\cal S}^{\mu\nu}_a &=& \frac{1}{w'_{A}} \{ 4(g^{\mu\nu} A^{(2)}_1 +P^{\mu}P^{\nu} A^{(2)}_2 + (P^{\mu}q^{\nu} + q^{\mu} P^{\nu}) A^{(2)}_3  \nonumber\\
&+& q^{\mu}q^{\nu} A^{(2)}_4) [M'^2+M''^2-q^2-(m'_1-m_2)^2 \nonumber\\
&-& (m''_1-m_2)^2 +(m'_1-m''_1)^2]-8 [g^{\mu\nu} A^{(2)}_1 Z_2  \nonumber\\
&+& q^{\mu}q^{\nu} (A^{(2)}_4 Z_2 + 2 \frac{q \cdot P}{q^2} A^{(1)}_2 A^{(2)}_1)] + 2(P^{\nu}A^{(1)}_{1} \nonumber\\
&+& q^{\nu} A^{(1)}_2) ( q^{\mu} [q^2- 2M'^2+N'_1 -N''_1 + 2(m'_1-m_2)^2 \nonumber\\
&-& (m'_1 -m''_1)^2] + P^{\mu} [q^2- N'_1 -N''_1 - (m'_1 -m''_1)^2]) \nonumber\\
&+& 4 q^{\mu}q^{\nu} [A^{(1)}_2 Z_2 + \frac{q\cdot P}{q^2} A^{(2)}_1]\} \ .
\eeqs
Finally, we obtain $ i {\cal A }^{\mu\nu} (a)$ as a function of the external four-momenta $P$ and $q$  with the following parameterization:
\beq
 i  {\cal A }^{\mu\nu} (a) = f^{a}_1(q^2) g^{\mu\nu}+  f^{a}_2(q^2) P^{\mu} q^{\nu} \ ,
\eeq
where the form factor that contributes to radiative decay width, $ f^{a}_1(q^2)
$, is given by
\begin{widetext}
\beqs
f^{a}_1 (q^2) &=&\frac{ee_{Q}N_c}{16\pi^3} \int \,
\frac{dx_2 d^2p'_{\perp}}{x_2 \hat N'_1 \hat N''_1}  h'_{A}h''_{P}\frac{4}{w'_A}\{ A^{(2)}_1 [M'^2 +M''^2- q^2 - (m'_1 -m_2)^2 -(m''_1 -m_2)^2 + (m'_1 -m''_1)^2]-2 A^{(2)}_1 Z_2  \} \nonumber\\
&=&\frac{ee_{Q}N_c}{16\pi^3} \int \,
\frac{dx_2 d^2p'_{\perp}}{x_2 \hat N'_1 \hat N''_1}  h'_{A}h''_{P}\frac{4}{w'_A} (-p'^2_{\perp}-\frac{(p'_{\perp}\cdot q_{\perp})^2}{q^2})   \{ [M'^2 +M''^2- q^2 - (m'_1 -m_2)^2 -(m''_1 -m_2)^2 \nonumber\\
&+& (m'_1 -m''_1)^2]- 2\hat N'_1- 2m'^2_1 +2 m^2_2 - 2(1-2x_1) M'^2- 2(q^2+ q\cdot P) \frac{p'_{\perp}\cdot q_{\perp}}{q^2}  \} \nonumber\\
&=& \frac{ee_{Q}}{16\pi^3} \int \,
\frac{dx_2 d^2p'_{\perp}}{x_1 M'_{0} M''_{0}} \varphi_p(p'_{\perp},x_2) \varphi(p''_{\perp},x_2)\left [-p'^2_{\perp}-\frac{(p'_{\perp}\cdot q_{\perp})^2}{q^2} \right] \nonumber\\
&\times& [(2x_1-1) M'^2 + M''^2+2x_1 M'^2_0 - q^2- 2(q^2+ q \cdot P) \frac{p'_{\perp}\cdot q_{\perp}}{q^2}] . \nonumber\\
\label{f1}
\eeqs
\end{widetext}

Similarly, for the right diagram in Fig.(\ref{p1}), we have the corresponding
amplitude:
\beq i {\cal A }^{\mu\nu} (b) = f^{b}_1(q^2) g^{\mu\nu}+
f^{b}_2(q^2) P^{\mu} q^{\nu} \ .
\eeq
This can be obtained from the result of the left-hand diagram with the
replacements $m'_1 \leftrightarrow m'_2$, $m''_1 \leftrightarrow m''_2$, $m_2
\leftrightarrow m_1$, $N_{e'_1} \leftrightarrow N_{e_2} $. The total form
factor $f_1 (q^2)$ is the sum of contribution from two diagrams:
\beq f_1 (q^2)
= f^{a}_1 (q^2)+f^{b}_1 (q^2) \ .
\label{f1sum}
\eeq
Taking the physical value $q^2 \to 0$ in the form factor, we can obtain
$|f_1(0)|^2$ and compute the decay width using Eq.(\ref{width}). The numerical
calculation of the decay width will be discussed in Section \ref{ANALYSIS}.


\section{Reduction to  non-relativistic limit in application to quarkonium systems}
\label{NR}

In this paper, we use the light-front formula discussed in Section \ref{LFQM} to
study the radiative decay (\ref{heta}). For this decay, the non-relativistic
electromagnetic dipole transition formulas are widely adopted
\cite{Brambilla:2004wf}. Thus it is interesting to investigate the consistency
between the LFQM and the non-relativistic dipole transition formulas in the
non-relativistic limit.  In this section we analyze the reduction of the
light-front formula for the decay width in the non-relativistic limit.  This
limit is relevant here because $(v/c)^2$ is substantially smaller than unity
for a heavy-quark $Q \bar Q$ state.  For a Coulombic potential, $\alpha_s =
v/c$, and current data give $\alpha_s = 0.21$ at a scale of $m_b = 4.5$ GeV,
yielding $(v/c)^2 = 0.04$ for the $\Upsilon$ system. There are several aspects
of the non-relativistic limit for the decay of a heavy quarkonium system:

\begin{enumerate}
 \item Masses of bound states.

   The masses of initial ($M'$) and final state ($M''$) are close to the
   sum their constituents, and the deviation is ${\cal O} (m^{-2}_Q)$
   corrections:
\beq \frac{ M'^2}{4m^2_Q } =1 + {\cal O} (m^{-2}_Q) \ , \
   \frac{ M''^2}{4m^2_Q } =1 + {\cal O} (m^{-2}_Q) \ .
\eeq
Here and below, by ${\cal O}(m^{-2}_Q)$ we mean ${\cal O}(|{\bf p}|^2/m_Q^2)$,
where ${\bf p}$ is a generic three-momentum in the parent meson rest frame.

   \item   No-recoil limit.

     In non-relativistic quantum mechanics, the final state after the E1
     radiative transition is assumed to carry approximately the same
     three-momentum as the initial state \cite{Daghighian:1987ru}. So the
     matrix element of this E1 transition is
\beq
   \langle {\bf r} \rangle \propto \langle f({\bf p}'') | {\bf r} | i ({\bf p}') \rangle, \quad  {\bf p}'' = {\bf p}' \ .
\eeq
In our analysis, we will adopt this no-recoil approximation.

  \item    Normalization of wavefunction.

    In non-relativistic quantum mechanics, the momentum-space wavefunction is
given by
\beq \langle {\bf p} | n,lm\rangle =R_{nl}(p) Y_{lm}(\theta,\phi)
\ ,
\eeq
with the normalization of the radial wavefunction
\beq
\int^{\infty}_0 dp \, p^2 R^*_{nl}(p)R_{nl}(p)=1 \ ,
\label{Norm}
\eeq
where here $p=|\bf p|$, and the normalization of the angular wavefunction
\beq
\int d\Omega \, Y^*_{lm}(\theta,\phi) Y_{l'm'}(\theta,\phi)=
\delta_{ll'}\delta_{mm'} \ .
\eeq
\end{enumerate}

In this paper we use harmonic oscillator wavefunctions for the quarkonium 1P
and 1S states. The general formula for harmonic oscillator wavefunctions in
momentum space that satisfy the usual quantum mechanics normalization in
Eq.(\ref{Norm}) is given by \cite{Godfrey:2015dia,Faiman:1968js}
\beqs
R_{nl}(p)&=&\frac{1}{\beta^{\frac{3}{2}}}\sqrt{\frac{2n}{\Gamma(n+l+\frac{1}{2})}}\nonumber\\
&\times&(\frac{p}{\beta})^{l}L^{l+\frac{1}{2}}_{n-1}(\frac{p^2}{\beta^2})\exp(-\frac{p^2}{2\beta^2})
\ ,
\eeqs
where $L^{l+\frac{1}{2}}_{n-1}({p^2}/{\beta^2})$ is an associated
Laguerre polynomial. Here, $\beta$ is a parameter with dimensions of momentum
that enters in the light-front wavefunction (\ref{psip}) (and should not be
confused with the dimensionless ratio $v/c$ that serves as a measure
of the non-relativistic property of a heavy-quark $Q \bar Q$ bound state.)
Specifically, for 1S and 1P states, we have
\beq
R_{\text{1S}}(p)=\frac{2}{\beta^{\frac{3}{2}}\pi^{\frac{1}{4}}}
\exp(-\frac{p^2}{2\beta^2}) \ ,
\eeq
and
\beq
R_{\text{1P}}(p)=  \sqrt{\frac{2}{3}}\frac{2}{\beta^{\frac{3}{2}}\pi^{\frac{1}{4}}} \exp(-\frac{p^2}{2\beta^2})\frac{p}{\beta} \ .
\label{1Pwave}
\eeq
Notice that the normalization of these wavefunctions is different from the
normalization of the light-front wavefunctions discussed in
Appendix \ref{wavefunction}. For example,
\beq
\psi (p) = \frac{1}{\sqrt{4\pi}}R_{\text{1S}}(p) \ .
\label{1Swave}
\eeq

In non-relativistic quantum mechanics, the width of an E1 decay
of the initial quarkonium state ${}^{1}P_{1}$ to the
final quarkonium state ${}^{1}S_{0} + \gamma $ is given by
\cite{Brambilla:2004wf}:
\beq
\Gamma({}^{1}P_{1} \to {}^{1}S_{0} + \gamma ) = \frac{4}{9}\alpha e^2_Q E^3_{\gamma}|I_3 (\text{1P} \to \text{1S})|^2
\label{NRE1}
\eeq
where $E_\gamma= |\bf q|$ is the energy of the emitted photon, and $I_3
(\text{1P} \to \text{1S})$ is the overlap integral in position space, which
represents the matrix element of the electric dipole operator:
\beq
I_3 (\text{1P} \to \text{1S})= \int^{\infty}_0 dr \,
r^3 R_{\text{1P}}(r)R^*_{\text{1S}}(r) \ .
\eeq
Similarly, we can define $I_5 (\text{1P} \to \text{1S})$, which appears in the
relativistic correction to the electric dipole transition width
\cite{Brambilla:2004wf}:
\beq I_5 (\text{1P} \to \text{1S})= \int^{\infty}_0 dr \,
r^5 R_{\text{1P}}(r)R^*_{\text{1S}}(r)
\eeq
For later use, we also list the analogous integrals in momentum space:
\beq
I^{p}_3 (\text{1P} \to \text{1S})= \int^{\infty}_0 dp \,
p^3 R_{\text{1P}}(p)R^*_{\text{1S}}(p) \ ,
\eeq
\beq
I^{p}_5 (\text{1P} \to \text{1S})= \int^{\infty}_0 dp \,
p^5 R_{\text{1P}}(p)R^*_{\text{1S}}(p) \ .
\eeq

We are now ready to reduce the light-front decay width in Eq.(\ref{width}) when
applied to quarkonium systems to the standard non-relativistic formula in
Eq. (\ref{NRE1}).  Using the explicit form in Eq.(\ref{f1}) and taking the
limit $q^2 \to 0$, the form factor in Eq.(\ref{f1sum}), we can write:
\begin{widetext}
\beqs
f_1 (q^2) &=&
 \frac{2ee_{Q}}{16 \pi^3}\int \frac{dx_2 d^2p'_{\perp}}{x_1 M'_{0} M''_{0}}   \varphi_p(p'_{\perp},x_2) \varphi(p''_{\perp},x_2)\left [-p'^2_{\perp}-\frac{(p'_{\perp}\cdot q_{\perp})^2}{q^2} \right]\cdot [(2x_1-1) M'^2 + M''^2+2x_1 M'^2_0 - 2( q \cdot P) \frac{p'_{\perp}\cdot q_{\perp}}{q^2}]  \nonumber\\
&=& -e e_{Q} \int \frac{dx_2 d^2p'_{\perp}}{x_1 M'_{0} M''_{0}}  \sqrt{\frac{dp'_z}{dx_2}}\sqrt{\frac{dp'_z}{dx_2}} \sqrt{\frac{e''_1 M'_0}{e'_1 M''_0}}  \psi_p(p'_{\perp},p'_z) \psi(p''_{\perp},p''_z) p'^2_{\perp} \nonumber\\
&\times& [(2x_1-1) M'^2 + M''^2+2x_1 M'^2_0 - 2( q \cdot P) \frac{p'_{\perp}\cdot q_{\perp}}{q^2}] \ ,
\label{f1zeroq}
\eeqs
where we use the explicit form of light-front momentum space wavefunction in
Appendix \ref{wavefunction}. This expression can be further simplified in the
no-recoil limit, which is a valid approximation in the study of an
electric dipole transition in the non-relativistic limit
\cite{Daghighian:1987ru}. In this
limit, we have
$\sqrt{\frac{e''_1 M'_0}{e'_1 M''_0}} \to 1 $, $M''_0 \to M'_0$ and
$\psi(p''^2_{\perp},p''^2_z) \to \psi(p'^2_{\perp},p'^2_z)$. The corrections
due to recoil effect are suppressed by powers of ($1/ m_Q$):
\beq
\sqrt{\frac{e''_1 M'_0}{e'_1 M''_0}} =\sqrt{\frac{2\sqrt{{\bf p}''^2+m^2_Q}}{\sqrt{{\bf p}'^2+m^2_Q}+\sqrt{{\bf p}''^2+m^2_Q}}} = 1- \frac{1}{8} \frac{({\bf p}'^2- {\bf p}''^2)^2}{m^4_Q}+ {\cal O} (m^{-6}_Q) \ ,
\eeq
\beq
M''_0= M'_0 + \frac{1}{2} \frac{({\bf p}''^2- {\bf p}'^2)}{m_Q} + {\cal O} (m^{-3}_Q) \ .
\eeq
The last term in Eq.(\ref{f1zeroq}), $- ( q \cdot P) \frac{p'_{\perp}\cdot
  q_{\perp}}{q^2}$, requires a more careful treatment. It seems that linear
$p'_{\perp}$ terms will not make contributions after integrating over
$p'_{\perp}$, but the Taylor expansion of the functions of $p''_{\perp}$ in the
integrands will generate a term that is proportional to $(p'_{\perp}\cdot
q_{\perp})$, and this can combine with $- ( q \cdot P) \frac{p'_{\perp}\cdot
q_{\perp}}{q^2}$ term to produce a $q^2$ independent term, which is non-zero
in the physical $q^2 \to 0$ and no-recoil limit. Firstly we should expand
$p''_{\perp}$ in powers of inverse of $ m_Q$:
\beq
p''_{\perp}=p'_{\perp}-x_2 q_{\perp}=  p'_{\perp}- q_{\perp} \left( \frac{1}{2}+ \frac{p'_z}{2\sqrt{m^2_Q+ p'^2_{\perp}+ p'^2_z}}\right)=p'_{\perp}- \frac{1}{2}q_{\perp} -\frac{1}{2}\frac{p'_z}{m_Q} q_{\perp}+  {\cal O}(m^{-2}_Q) \ .
\eeq
We find in the physical limit $q^2 \to 0$, the dominant contribution to the
$(p'_{\perp}\cdot q_{\perp})$ term comes from the expansion of
$\psi(p''_{\perp},p''_z)$. Since $\psi(p''_{\perp},p''_z)$ is the wavefunction
of the 1S state, it is a function of ${\bf p}''^2$.  Hence, we can write
$\psi(p''_{\perp},p''_z)=\psi(p''^2_{\perp},p''^2_z)$ and expand it as follows:
\beq \psi(p''^2_{\perp},p''^2_z) \approx \psi(p'^2_{\perp},p'^2_z) - p'_{\perp}
\cdot q_{\perp} [\frac{d\psi}{dp'^2_{\perp}}] + {\cal O}(m^{-2}_Q) =
\psi(p'^2_{\perp},p'^2_z) + p'_{\perp} \cdot q_{\perp} \frac{1}{2\beta^2}
\psi(p'^2_{\perp},p'^2_z) + {\cal O}(m^{-2}_Q) \ ,
\eeq
where we use the
explicit form of $\psi(p''_{\perp},p''_z) \propto \exp[-{{\bf
    p}''}^2/(2\beta^2)]$ to calculate its derivative.  Plugging the
expansion of $\psi(p''_{\perp},p''_z)$ into the integrands, we find the
contribution of the
$- (q \cdot P) \frac{p'_{\perp}\cdot q_{\perp}}{q^2}$ terms is
\beq \int ...
\psi(p''_{\perp},p''_z) [ - ( q \cdot P) \frac{p'_{\perp}\cdot q_{\perp}}{q^2}]
|_{q^2 \to 0}= - ( q \cdot P) \int ... [\frac{d\psi}{dp'^2_{\perp}}] [
\frac{(p'_{\perp}\cdot q_{\perp})^2}{q^2_{\perp}}]= \frac{1}{2\beta^2} ( q
\cdot P) \int ...  \psi(p'_{\perp},p'_z) \frac{1}{2}p'^2_{\perp} \ .
\eeq
After this calculation, in the physical $q^2=0$ and no-recoil limit, the
form factor $f_1 (q^2\to 0)$ is given by
\beqs f_1 ( 0)
&\approx& -ee_{Q} \int \frac{dp'_z d^2p'_{\perp}}{x_1 M'^2_{0} }   \psi_p(p'_{\perp},p'_z) \psi(p'_{\perp},p'_z) p'^2_{\perp} \cdot [(2x_1-1) M'^2 + M''^2+2x_1 M'^2_0 + 2( q \cdot P) \frac{1}{4\beta^2}p'^2_{\perp}]  \nonumber\\
&=&-ee_{Q} \int {dp'_z d^2p'_{\perp}}  \psi_p(p'_{\perp},p'_z) \psi(p'_{\perp},p'_z) p'^2_{\perp} \cdot \left [2+ \frac{2 M'^2}{4(m^2_Q+ p'^2_{\perp}+p'^2_z)}\right. \nonumber\\
&+& \left. \frac{M''^2-M'^2}{2(m^2_Q+ p'^2_{\perp}+p'^2_z)-2p'_z\sqrt{m^2_Q+
      p'^2_{\perp}+p'^2_z}} + \frac{2|{\bf q}|M'}{(m^2_Q+
    p'^2_{\perp}+p'^2_z)-p'_z\sqrt{m^2_Q+ p'^2_{\perp}+p'^2_z}}
  \frac{1}{4\beta^2}p'^2_{\perp}\right] ,
\eeqs
where we use the kinematic
relation $(q\cdot P)= 2 |{\bf q}| M'$.  In the non-relativistic limit, it is
more convenient to use notation of wavefunctions in non-relativistic quantum
mechanics. Using Eq. (\ref{1Swave}), $f_1 (q^2\to 0)$ can be rewritten as
\beqs
f_1 (q^2 \to 0)
&\approx& -\frac{1}{4\pi}  \cdot {\frac{\sqrt{2}}{\beta}}\cdot ee_{Q} \int d^3
{\bf p}' \,
 R_{\text{1S}}({\bf p}') R_{\text{1S}}({\bf p}') p'^2_{\perp} \cdot \left [2+ \frac{ 2M'^2}{4(m^2_Q+  {\bf p}'^2)}\right. \nonumber\\
&+& \left. \frac{M''^2-M'^2}{2(m^2_Q+ {\bf p}'^2)-2p'_z\sqrt{m^2_Q+ {\bf p}'^2}} +
  \frac{2|{\bf q}|M'}{(m^2_Q+{\bf p}'^2)-p'_z\sqrt{m^2_Q+ {\bf p}'^2}}
  \frac{1}{4\beta^2}p'^2_{\perp}\right] \ .
\label{f1norecoil}
\eeqs
\end{widetext}


\subsection{Leading-order non-relativistic approximation}

In the leading-order non-relativistic approximation, we neglect the ${\cal
  O}(m^{-2}_Q)$ contributions in Eq.(\ref{f1norecoil}), so $f_1 (q^2 \to 0)$ is
given by
\begin{widetext}
\beqs
f_1 (q^2 \to 0)
&\approx& - 4\pi  \cdot {\frac{\sqrt{2}}{\beta}}\cdot e e_{Q}\int d^3 {\bf p}'
\,  R_{\text{1S}}({\bf p}') R_{\text{1S}}({\bf p}') p'^2_{\perp} \cdot \left [2+ \frac{ 2M'^2}{4m^2_Q }+ \frac{M''^2-M'^2}{2m^2_Q} +\frac{2|{\bf q}|M'}{m^2_Q} \frac{1}{4\beta^2}p'^2_{\perp}+ {\cal O}(m^{-4}_Q)\right]\nonumber\\
&=& - \frac{1}{4\pi}  \cdot {\frac{\sqrt{2}}{\beta}}\cdot ee_{Q}
 \int d^3 {\bf p}' \,
 R_{\text{1S}}({\bf p}') R_{\text{1S}}({\bf p}') p'^2_{\perp} \cdot 4 + {\cal O}(m^{-2}_Q)
\label{LO1}
\eeqs
\end{widetext}
This integral can be simplified by using symmetric property of functions in the
integrands. For function $F({\bf p}^2)$ that have spherical symmetry, the
following relation is satisfied:
\beq \int d^3 {\bf p} \, F({\bf p}^2)p_i p_j =
\frac{1}{3} \delta_{ij} \int d^3 {\bf p} \, F({\bf p}^2) {\bf p}^2 \ .
\eeq
So Eq.(\ref{LO1}) can be written as
\beqs
f_1 ( 0)&=&-\frac{4}{4\pi}  \cdot {\frac{\sqrt{2}}{\beta}}\cdot ee_{Q}
\int d^3 {\bf p}' \, R_{\text{1S}}({\bf p}') R_{\text{1S}}({\bf p}') p'^2_{\perp} \nonumber\\
&=&-\frac{2}{3}\cdot\frac{4}{4\pi}   \cdot {\frac{\sqrt{2}}{\beta}}\cdot ee_{Q}
\int d^3 {\bf p}' \,
  R_{\text{1S}}({\bf p}') R_{\text{1S}}({\bf p}') {\bf p}'^2 \nonumber\\
&=&-\frac{2}{3}\cdot {4}   \cdot {\frac{\sqrt{2}}{\beta}}\cdot ee_{Q}
\int^{\infty}_{0} d p' \, { p}'^4  R_{\text{1S}}({ p}') R_{\text{1S}}({ p}') \ ,  \nonumber\\
\eeqs
where $p'$ denotes the radial coordinate in the three dimensional
momentum space (and should not be confused with a four-momentum). Using the
definition of wavefunction in Eq.(\ref{1Pwave}),
\beq R_{\text{1P}}({ p}')=
\frac{1}{\beta}\sqrt{\frac{2}{3}} R_{\text{1S}}({ p}')p' \ ,
\eeq
we find that this integral is proportional to $I^{{p}}_3(\text{1P} \to
\text{1S}) $:
\beqs
f_1 ( 0)
&=&-\frac{2}{3}\cdot {4}   \cdot {\frac{\sqrt{2}}{\beta}}\cdot ee_{Q}
\int^{\infty}_{0} d p' \,
{ p}'^4  R_{\text{1S}}({ p}') R_{\text{1S}}({ p}')  \nonumber\\
&=&-\sqrt{\frac{3}{2}}\cdot\frac{2}{3}\cdot {4}   \cdot \sqrt{2}\cdot ee_{Q}
\int^{\infty}_{0} d p' \, { p}'^3  R_{\text{1P}}({ p}') R_{\text{1S}}({ p}')  \nonumber\\
&=&-\sqrt{\frac{3}{2}}\cdot\frac{2}{3}\cdot {4}   \cdot \sqrt{2}\cdot ee_{Q} \cdot I^{{p}}_3(\text{1P} \to \text{1S})  \ .
\eeqs
Now $I^{{p}}_3(\text{1P} \to \text{1S}) $ is proportional to
$I_3(\text{1P} \to \text{1S}) $, which is evident in non-relativistic quantum
mechanics, where we have the operator relation:
\beq
  \frac{{\bf p}}{m}= i [H,{\bf r}] \ ,
\eeq
so that
\beq
|\langle f |  \frac{{\bf p}}{m} | i \rangle | = |\langle f  [H,{\bf r}]  | i \rangle |  = (E_i-E_f) |\langle f |  {\bf r}  | i \rangle | \ .
\eeq
In the non-relativistic limit, the mass can be interpreted as the reduced mass
of the $\bar Q Q$ two-body system $m=\mu'= m_Q/2$, and in
non-relativistic quantum mechanics the photon energy is the difference of
energy levels between initial and final state, $E_i-E_f \approx |{\bf q}|$, so
we have
\beq
I^{{p}}_3(\text{1P} \to \text{1S}) = |{\bf q}| \mu'   \cdot   I_3(\text{1P} \to \text{1S}) \ .
\eeq
Then $f_1 ( 0)$ can be expressed as
\beqs
f_1 ( 0)&=&-\sqrt{\frac{3}{2}}\cdot\frac{2}{3}\cdot {4}   \cdot \sqrt{2}\cdot ee_{Q}\cdot I_3(\text{1P} \to \text{1S})\cdot |{\bf q}|\mu' \ .  \nonumber\\
\eeqs
Plugging this expression of $f_1 ( 0)$ into the formula for the decay width in
Eq.(\ref{width}), we get radiative decay width of $A({}^{1}P_{1}) \to
P({}^1S_{0}) +\gamma $ in the leading order non-relativistic and no-recoil
approximation:
\beqs
\Gamma_{\text{NR}}&=&\frac{| {\bf q}|^3\mu'^2}{12 \pi M'^2} \frac{3}{2}\cdot\frac{4}{9}\cdot {16}   \cdot 2\cdot e^2 e^2_{Q} | I_3(\text{1P} \to \text{1S})|^2\nonumber\\
&=& \left[ \frac{16\mu'^2}{M'^2}\right ] \cdot \frac{4}{9}\cdot \alpha e^2_{Q} |{\bf q}|^3 \cdot | I_3(\text{1P} \to \text{1S})|^2 \nonumber\\
&=&  \frac{4}{9}\cdot \alpha  e^2_{Q} |{\bf q}|^3 \cdot | I_3(\text{1P} \to \text{1S})|^2  \cdot (1+{\cal O}( m^{-2}_Q))\nonumber\\
&\approx& \frac{4}{9}\cdot \alpha  e^2_{Q} |{\bf q}|^3 \cdot |I_3(\text{1P} \to \text{1S})|^2 \ ,
\label{NRwidth}
\eeqs
where we have made use of the approximate relations of masses:
\beq
\mu' = \frac{m_Q}{2} \ , M' \simeq 2m_Q\ , \ \to 
\left[ \frac{16\mu'^2}{M'^2}\right ]  \simeq 1 \ .
\eeq
Eq.(\ref{NRwidth}) matches the non-relativistic electric dipole transition
formula for transition ${}^{1}P_{1}$ $\to$ ${}^1S_{0}$ in Eq.(\ref{NRE1}),
which proves the validity of light-front framework in the non-relativistic
limit in the application to heavy quarkonium systems.


\subsection{Next-to-leading order correction}

We next include the ${\cal O}(m^{-2}_Q)$ contributions in Eq.(\ref{f1norecoil})
with the no-recoil approximation. In this case, $f_1 (q^2 \to 0)$ is given by
\begin{widetext}
\beqs
f_1 (0)
&\approx& -\frac{1}{4\pi}  \cdot {\frac{\sqrt{2}}{\beta}}\cdot ee_{Q}
\int d^3 {\bf p}'\,
  R_{\text{1S}}({\bf p}') R_{\text{1S}}({\bf p}') p'^2_{\perp} \cdot \left [2 + \frac{2 M'^2}{4m^2_Q}(1-\frac{ {\bf p}'^2}{m^2_Q}) +  \frac{M''^2-M'^2}{2m^2_Q}  + \frac{2|{\bf q}|M'}{m^2_Q} \frac{1}{4\beta^2}p'^2_{\perp}\right] \nonumber\\
&=&
 -\frac{1}{4\pi}  \cdot {\frac{\sqrt{2}}{\beta}}\cdot ee_{Q}
\int d^3 {\bf p}' \,
 R_{\text{1S}}({\bf p}') R_{\text{1S}}({\bf p}') p'^2_{\perp} \cdot 4\left [1-\frac{1}{2}\frac{ {\bf p}'^2}{m^2_Q} + \frac{|{\bf q}|}{m_Q} \frac{1}{4\beta^2}p'^2_{\perp}+R_{\text{1P,1S}} + {\cal O}(m^{-4}_Q)\right]  \ , \nonumber\\
 &=&
 -\frac{1}{4\pi}  \cdot {\frac{\sqrt{2}}{\beta}}\cdot ee_{Q}
\int d^3 {\bf p}' \,
 R_{\text{1S}}({\bf p}') R_{\text{1S}}({\bf p}')\cdot 4\left [\frac{2}{3} \cdot (1+R_{\text{1P,1S}}){\bf p}'^2-\frac{1}{3}\frac{ {\bf p}'^4}{m^2_Q}  + \frac{2}{15} \frac{|{\bf q}|}{ m_Q} \frac{1}{\beta^2}{\bf p}'^4 + {\cal O}(m^{-4}_Q)\right]  \ ,\nonumber\\
 &=&
 -  {\sqrt{3}}\cdot ee_{Q}\cdot 4\cdot \frac{2}{3} \cdot I^{{p}}_3(\text{1P}\to \text{1S}) \left [  1+R_{\text{1P,1S}}-\left ( \frac{1}{2}\frac{ 1}{m^2_Q}  - \frac{1}{5} \frac{|{\bf q}|}{ m_Q} \frac{1}{\beta^2} \right) \frac{I^{{p}}_5(\text{1P}\to \text{1S})}{I^{{p}}_3(\text{1P}\to \text{1S})}+ {\cal O}(m^{-4}_Q)\right]  \ ,\nonumber\\
&=&
-  {\sqrt{3}}\cdot ee_{Q}\cdot 4\cdot \frac{2}{3} \cdot |{ \bf q}|\mu'\cdot I_3(\text{1P}\to \text{1S}) \left [  1+R_{\text{1P,1S}}-|{\bf q}|^2\left ( \frac{1}{2}\frac{\mu'^2}{m^2_Q}  - \frac{1}{5} \frac{\mu'}{ m_Q}  \right) \frac{I_5(\text{1P}\to \text{1S})}{I_3(\text{1P}\to \text{1S})}+ {\cal O}(m^{-4}_Q)\right]  \ ,\nonumber\\
 \nonumber\\
\label{f1norecoilNLO}
\eeqs
\end{widetext}
where we have made use of the symmetry property of integral for the function
$F({\bf p}^2)$ that has spherical symmetry:
\beqs
&&\int d^3 {\bf p}\, F({\bf p}^2)p_i p_j p_k p_l  \nonumber\\
&&= \frac{1}{15} (\delta_{ij}\delta_{kl}+
\delta_{ik}\delta_{jl}+\delta_{il}\delta_{jk}) \int d^3 {\bf p} \,
F({\bf p}^2) {\bf p}^4 \ , \nonumber\\
\eeqs
and $R_{\text{1P,1S}}$ is given by
\beq R_{\text{1P,1S}}=
\frac{M''^2-M'^2}{8m^2_Q}+ \frac{M'^2-4m^2_Q}{8m^2_Q} \sim {\cal O}(m^{-2}_Q)
\ .
\eeq
Combining Eq.(\ref{f1norecoilNLO}) and Eq.(\ref{width}), we obtain the
next-to-leading order (${\cal O}(m^{-2}_Q)$) formula for the radiative decay
width for the heavy quarkonium systems (${}^{1}P_{1} \to {}^{1}S_{0}$) in the
non-relativistic and no-recoil approximation:
\beqs
\Gamma_{\text{NLO}}&=&\Gamma_{\text{NR}}[1+R_{\text{1P,1S}}-{|{\bf q}|^2}\left (\frac{1}{2}\frac{\mu'^2}{m^2_Q} - \frac{1}{5} \frac{\mu'}{ m_Q} \right) \nonumber\\
&\times& \frac{I_5(\text{1P}\to \text{1S})}{I_3(\text{1P}\to \text{1S})}+ {\cal
  O}(m^{-4}_Q)]^2 \ .
\eeqs
%


\section{Analysis of radiative transitions of $ h_{c}(1P) $ and
$h_{b}(1P)$ }
\label{ANALYSIS}

\begin{widetext}

\begin{table}
\begin{ruledtabular}
  \caption{Decay width (in units of keV) of $h_{c}(1P) \to \eta_{c}(1S)+\gamma$
    in the light-front quark model, denoted LFQM, as compared with experimental
    data from \cite{PDG}, denoted exp.(PDG) and predictions from other
    theoretical models, including non-relativistic potential model
    (NR)\cite{Brambilla:2004wf,Barnes:2005pb,Voloshin:2007dx}, relativistic
    quark model (R)\cite{Ebert:2002pp}, the Godfrey-Isgur potential model
    (GI)\cite{Barnes:2005pb}, screened potential models with zeroth-order
    wavefunctions (${\text{SNR}}_{0}$) and first-order relativistically
    corrected wavefunctions (${\text{SNR}}_{1}$) \cite{Li:2009zu}. For
    experimental data, we use the PDG value of the total width
    $\Gamma_{h_{c}(1P)}=700 \pm280 \pm 220$ keV and $BR(h_{c}(1P) \to
    \eta_{c}(1S)+\gamma) = 51 \pm 6$ \% \cite{PDG}.  } \label{tabhc}
\begin{tabular}{cccccccc}
  Decay mode & LFQM &  exp.(PDG)\cite{PDG}    & NR\cite{Brambilla:2004wf} & R\cite{Ebert:2002pp}   & NR/GI\cite{Barnes:2005pb} & \cite{Voloshin:2007dx} & ${\text{SNR}}_{0/1}$\cite{Li:2009zu}\\
  \hline
$h_{c}(1P) \to \eta_{c}(1S)+\gamma$ & $398 \pm 99$ & $ 357 \pm 280$   & 482 & 560  & 498 / 352  &  650  & 764/323\\
\end{tabular}
\end{ruledtabular}
\end{table}

\begin{table}
\begin{ruledtabular}
  \caption{Decay width (in units of keV) of $h_{b}(1P) \to \eta_{b}(1S)+\gamma$
    in the light-front quark model, denoted LFQM, as compared with predictions
    from other theoretical models, including non-relativistic potential model
    (NR)\cite{Brambilla:2004wf}, relativistic quark model
    (R)\cite{Ebert:2002pp}, the Godfrey-Isgur potential model
    (GI)\cite{Godfrey:2015dia}, screened potential models with zeroth-order
    wavefunctions (${\text{SNR}}_{0}$) and first-order relativistically
    corrected wavefunctions (${\text{SNR}}_{1}$) \cite{Li:2009nr} and the
    nonrelativistic constituent quark model (CQM)\cite{Segovia:2016xqb}.
  } \label{tabhb}
\begin{tabular}{cccccccc}
  Decay mode & LFQM & NR\cite{Brambilla:2004wf} &  R\cite{Ebert:2002pp}   & ${{\text {SNR}}_{0/1}}$\cite{Li:2009nr}  & GI\cite{Godfrey:2015dia} & CQM\cite{Segovia:2016xqb} \\
  \hline
$h_{b}(1P) \to \eta_{b}(1S)+\gamma$ & $37.5 \pm 7.5$ & 27.8&
 52.6& 55.8/ 36.3  & 35.7  & 43.7\\
\end{tabular}
\end{ruledtabular}
\end{table}

\end{widetext}

\begin{figure}
\begin{center}
\resizebox{0.45\textwidth}{!}{%
\includegraphics{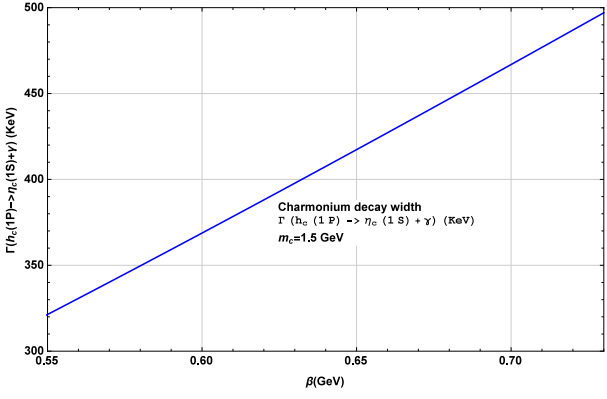}
}%
\caption {Decay width for $h_{c}(1P) \to \eta_{c}(1S) + \gamma$ (KeV) as a function of $\beta_{h_c(1P)(\eta_{c})(1S)}$
  in LFQM, with $m_c=1.5$ GeV.}
\label{cbeta}
\end{center}
\end{figure}

\begin{figure}
\begin{center}
\resizebox{0.45\textwidth}{!}{%
\includegraphics{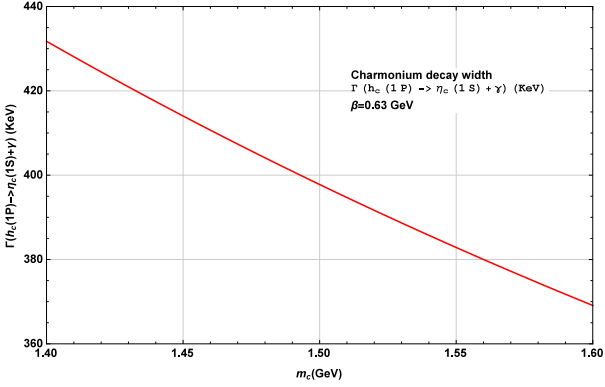}
}%
\caption {Decay width for $h_{c}(1P) \to \eta_{c}(1S) + \gamma$ (KeV) as a
  function of  $m_c$ in the LFQM, with $\beta_{h_c(1P)(\eta_{c}(1S))}= 0.63$ GeV.}
\label{cmass}
\end{center}
\end{figure}

\begin{figure}
\begin{center}
\resizebox{0.45\textwidth}{!}{%
\includegraphics{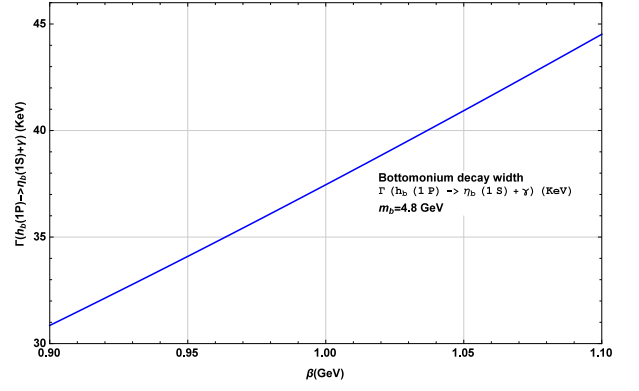}
}%
\caption {Decay width for $h_{b}(1P) \to \eta_{b}(1S) + \gamma$ (KeV) as a function of $\beta_{h_b(1P)(\eta_{b}(1S))}$
  in the LFQM, with $m_b=4.8$ GeV.}
\label{bbeta}
\end{center}
\end{figure}

\begin{figure}
\begin{center}
\resizebox{0.45\textwidth}{!}{%
\includegraphics{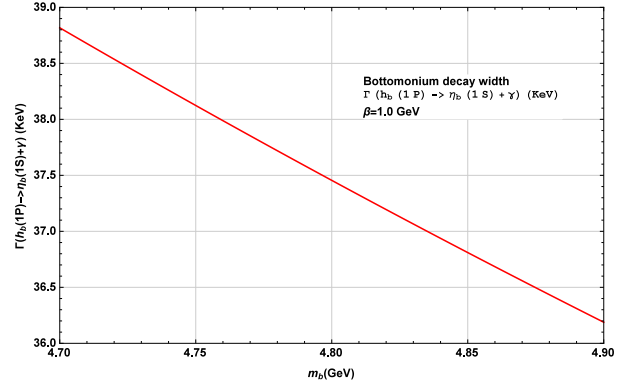}
}%
\caption {Decay width for $h_{b}(1P) \to \eta_{b}(1S) + \gamma$ (KeV) as a function of $m_b$
  in the LFQM, with $\beta_{h_b(1P)(\eta_{b}(1S))}= 1.0$ GeV.}
\label{bmass}
\end{center}
\end{figure}

In this section we apply the radiative transition formulas for the decay
$1^{+-} ({}^{1}P_1) \to 0^{-+}({}^{1}S_{0}) +\gamma$ in the framework of the
light-front quark model, which we reviewed in Section \ref{LFQM}, to study the
radiative decay of the $c \bar c$ state $h_{c1}(1P) $ via the channel
$h_{c}(1P) \to \eta_{c}(1S)+\gamma$ and the $b \bar b$ state $h_{b}(1P)$ via
the channel $ h_{b}(1P) \to \eta_{b}(1S)+\gamma$.  We present the results of
our numerical calculations of decay widths. Our results extend those which we
previously presented with Ke and Li in \cite{Ke:2013zs}.
For the charmonium $h_{c}(1P)$ radiative decay, we compare our result
with experimental data on the width, as listed in the Particle Data Group
Review of Particle Properties (RPP) \cite{PDG}. We also list the theoretical
calculations from other models, including non-relativistic potential model
(NR)\cite{Brambilla:2004wf,Barnes:2005pb,Voloshin:2007dx}, relativistic quark
model (R)\cite{Ebert:2002pp}, the Godfrey-Isgur potential model
(GI)\cite{Barnes:2005pb}, screened potential models with zeroth-order
wavefunctions (${\text{SNR}}_{0}$) and first-order relativistically corrected
wavefunctions (${\text{SNR}}_{1}$) \cite{Li:2009zu}.

Although the PDG lists the width for the decay $h_{c}(1P) \to \eta_{c}(1S) +
\gamma$, it does not list the width for the $h_{b}(1P) \to \eta_{b}(1S) +
\gamma$ decay, only the branching ratio.  Since our calculation yields the
width itself, and a calculation of the branching ratio requires division by the
total width, we therefore compare our results on the widths for these
decays with predictions from other models, including the non-relativistic
potential model (NR)\cite{Brambilla:2004wf}, the relativistic quark model
(R)\cite{Ebert:2002pp}, the Godfrey-Isgur potential model
(GI)\cite{Godfrey:2015dia}, screened potential models with zeroth-order wave
functions (${\text{SNR}}_{0}$) and first-order relativistically corrected wave
functions(${\text{SNR}}_{1}$) \cite{Li:2009nr}, as well as the nonrelativistic
constituent quark model (CQM)\cite{Segovia:2016xqb}.

First, we study the radiative decay $h_{c}(1P) \to \eta_{c}(1S) + \gamma$ in
the LFQM, which depends on the corresponding harmonic oscillator wavefunction
($\beta_{h_{c}(\eta_{c})}$) and the effective charm quark mass, $m_c $. For the
central values of $m_c$ and the wavefunction parameters $\beta$, we use the
central values of these parameters suggested by previous study of LFQM
\cite{Shi:2016nxr}:
\beq
m_c = 1.5 \pm 0.1  \ \text{GeV} \ .
\label{mc}
\eeq
\beq
\beta_{h_{c}(\eta_{c})}= 0.63 \pm 0.1 \ \text{GeV}  \ .
\label{betahc}
\eeq
While Ref. \cite{Ke:2013zs} allowed a 10 \% variation in input parameters, we
investigate a somewhat larger variation, as indicated in Eqs. (\ref{mc}) and
(\ref{betahc}).  We present our numerical results in Table \ref{tabhc}, with
the uncertainties arising from the uncertainties in the $\beta$ parameters and
the value of $m_c$. We also plot the predicted width as a function of the input
values for the charm quark mass $m_c$ and wavefunction structure parameter
$\beta_{h_c(\eta_c)}$ in Fig. \ref{cbeta} and Fig. \ref{cmass}. From these
results, we find that the main theoretical uncertainties come from
variation of $\beta_{h_c(\eta_c)}$.  With the same central value for
$\beta_{h_c(\eta_c)}$ as was used in \cite{Ke:2013zs}, we obtain a somewhat
smaller central value for the width, namely 398 keV as contrasted with 685 keV
in \cite{Ke:2013zs}. As is evident from our Table \ref{tabhc}, our current
result for this width agrees well with experimental data within the range of
experimental and theoretical uncertainties. The experimental data have
substantial uncertainties, and our result is relatively close to the central
experimental value, compared to other non-relativistic models.

Next we study radiative decay of $h_{b}(1P) \to \eta_{b}(1S) + \gamma$ in
LFQM. For the central value of the effective bottom/beauty quark mass $m_b$, we
use the value suggested by the previous LFQM study \cite{Ke:2013zs} (see also
\cite{Shi:2016nxr}). For the wavefunction parameter
$\beta_{h_{b}(1P)(\eta_{b}(1S))}$, we estimate this to be in the range $\beta
\sim 0.9 - 1.3 \ \text{GeV}$, which is suggested in \cite{Godfrey:2015dia},
where $\beta$ is fitted by equating the rms radius of the harmonic oscillator
wavefunction for the specified states with the rms radius of the wavefunctions
calculated using the relativized quark model.  Our values for these input
parameters are:
\beq
m_b = 4.8 \pm 0.1  \ \text{GeV} \ .
\label{mb}
\eeq
\beq
\beta_{h_{b}(1P)(\eta_{b}(1S))}= 1.0 \pm 0.1 \ \text{GeV} \ .
\label{betahb}
\eeq
We list the numerical results in the LFQM in Table \ref{tabhb}. For comparison,
we also list other theoretical calculations from various types of models,
including the non-relativistic potential model (NR)\cite{Brambilla:2004wf}, the
relativistic quark model (R)\cite{Ebert:2002pp}, the Godfrey-Isgur potential
model (GI)\cite{Godfrey:2015dia}, screened potential models with zeroth-order
wavefunctions (${\text{SNR}}_{0}$) and first-order relativistically corrected
wavefunctions(${\text{SNR}}_{1}$) \cite{Li:2009nr} and the non-relativistic
constituent quark model (CQM)\cite{Segovia:2016xqb}. As can be seen from Table
\ref{tabhb}, with the given range of uncertainties, our value agrees with
predictions from the non-relativistic potential model
(NR)\cite{Brambilla:2004wf}, the Godfrey-Isgur potential model
(GI)\cite{Godfrey:2015dia} and screened potential models with relativistically
corrected wavefunctions (${\text{SNR}}_{1}$) \cite{Li:2009nr}.  To show the
theoretical uncertainties arising from uncertainties in the
$\beta_{h_{b}(1P)(\eta_{b}(1S))}$ parameter and the value of $m_b$, we also
plot the decay width for $h_{b}(1P) \to \eta_{b}(1S) + \gamma$ as a function of
these parameters in Fig. \ref{bbeta} and Fig. \ref{bmass}. We find that the
width is not very sensitive to the variation of $m_b$ and the main
uncertainties arise from the uncertainty in the wavefunction parameter
$\beta_{h_{b}(1P)(\eta_{b}(1S))}$.

These results show that the light-front quark model with phenomenological meson
wavefunctions (specifically, harmonic oscillator wavefunctions)
is suitable for the calculation of quarkonium ${}^{1}P_{1} \to
{}^{1}S_{0} + \gamma$ radiative decay widths, since this model gives reasonable
predictions for these widths, as compared with experimental data and other
theoretical models.


\section{Conclusion}
\label{CON}

In this paper we have revisited the calculation of the radiative decay width of
a $1^{+-}$ axial vector meson A to a $0^{-+}$ pseudoscalar meson P via the
channel $1^{+-} \to 0^{-+} +\gamma$ in the LFQM approach, extending our
previous work in Ref. \cite{Ke:2013zs}. As part of our analysis, we have
presented the reduction of the LFQM results in the non-relativistic limit and
have shown the connection with the non-relativistic electric dipole
transition formula for heavy quarkonium systems. We have then applied the LFQM
formula to the radiative decays $h_{c}(1P) \to \eta_{c}(1S) + \gamma$ and
$h_{b}(1P) \to \eta_{b}(1S) + \gamma$. We have performed numerical calculations
and have compared our results with experimental data and other model
predictions.  We have shown that our results are in reasonable agreement with
data and other model calculations.


\begin{acknowledgments}
  We are grateful to Prof. Robert Shrock for his illuminating suggestions and
  assistance. This research was partially supported by the NSF grant
  NSF-PHY-13-16617. We would like to thank Profs. Hong-Wei Ke and Xue-Qian Li
  for collaboration on our previous related work \cite{Ke:2013zs}.
\end{acknowledgments}


\begin{appendix}


\section{The wavefunctions}
\label{wavefunction}

The normalization of the S-wave meson wavefunction in the light-front framework
is
\beq
\frac{1}{2(2\pi)^3} \int dx_2 dp^2_{\perp} \, |\varphi(x_2, p_{\perp})|^2 =1.
\eeq
Here $\varphi(x_2, p_{\perp})$ is related to the wavefunction in normal
coordinates $\psi(p)$ by
\beq
\varphi (x_2, p_{\perp})= 4\pi^{\frac{3}{2}} \sqrt{\frac{d p_z}{dx_2}}\psi (p) \ , \quad \frac{dp_z}{dx_2}= \frac{e'_1e_2}{x_1x_2 M'_{0}} \
\eeq
The normalization of $\psi(p)$ is given by
\beq
\int d{\bf  p}^3 \, |\psi(p)|^2 = 4\pi  \int  p^2 dp \, |\psi(p)|^2 =1 \ .
\eeq
The normalization for the P-wave meson wavefunction in the light-front
framework is \cite{Cheng:2003sm}
\beq
\frac{1}{2(2\pi)^3} \int dx_2 dp^2_{\perp} \,
|\varphi_{p}(x_2, p_{\perp})|^2 p_i p_j=\delta_{ij} \ ,
\label{normalP}
\eeq
where $p_i=(p_x,p_y,p_z)$. In terms of the P-wave wavefunction in normal
coordinates,
\beq
\varphi_{p} (x_2, p_{\perp})= 4\pi^{\frac{3}{2}} \sqrt{\frac{d p_z}{dx_2}}\psi_p (p) \ , \quad \frac{dp_z}{dx_2}= \frac{e'_1e_2}{x_1x_2 M'_{0}} \ .
\eeq
we have the following normalization condition:
\beq
\frac{1}{3} \cdot 4\pi  \int^{\infty}_{0}|\psi_{p}(p)|^2 p^4 dp = 1 \ .
\eeq

For the gaussian type 1P and 1S wavefunctions, we have the relation
\beq
\psi_{p}(p)=\sqrt{\frac{2}{\beta^2}} \psi (p) \ .
\label{psip}
\eeq
The explicit form of 1-S harmonic oscillator wavefunction in the light-front
approach is given by \cite{Cheng:2003sm}
\beqs
\psi(p) &=& \left( \frac{1}{\beta^2 \pi}  \right)^{\frac{3}{4}} \exp \left(  -\frac{1}{2}\frac{p^2}{\beta^2} \right) \ .
\eeqs
%


\section{Some expressions in the light-front formalism}
\label{expression}

In the covariant light-front formalism we have
\beqs
M'^2_0 &=&(e'_1 + e_2)^2 = \frac{p'^2_{\perp}+m'^2_1}{x_1}+ \frac{p'^2_{\perp}+m^2_2}{x_2} \nonumber\\
M''^2_0 &=&(e''_1 + e_2)^2 = \frac{p''^2_{\perp}+m''^2_1}{x_1}+ \frac{p''^2_{\perp}+m^2_2}{x_2}  \nonumber\\
\tilde M'_{0} &=&  \sqrt{M'^2_0-(m'_1-m_2)^2}\nonumber\\
\tilde M''_{0} &=& \sqrt{M''^2_0-(m''_1-m_2)^2} \nonumber\\
p'_z &=& \frac{x_2 M'_{0}}{2}- \frac{m^2_2+p'^2_{\perp}}{2x_2 M'_{0}} \nonumber\\
p''_z &=& \frac{x_2 M''_{0}}{2}- \frac{m^2_2+p''^2_{\perp}}{2x_2 M''_{0}} .
\eeqs

The explicit expressions for $A^{(i)}_{j} (i,j=1 \sim 4) $ and $Z_2$ are
\beqs
&&A^{(1)}_1=\frac{x_1}{2} \ , \ A^{(1)}_2= A^{(1)}_1- \frac{p'_{\perp}\cdot q_{\perp}}{q^2}\ , \nonumber\\
&&A^{(2)}_{1}=-p'^2_{\perp}-\frac{(p'_{\perp}\cdot q_{\perp})^2}{q^2}, \ A^{(2)}_2= (A^{(1)}_1)^2, \nonumber\\
&& A^{(2)}_3= A^{(1)}_1 A^{(1)}_2 \ , \   A^{(2)}_4= (A^{(1)}_2)^2-\frac{1}{q^2}A^{(2)}_1 \ ,\nonumber\\
&& A^{(3)}_1 = A^{(1)}_1 A^{(2)}_1 \ , A^{(3)}_2 = A^{(1)}_2 A^{(2)}_1 \ , \nonumber\\
&& A^{(3)}_3 = A^{(1)}_1 A^{(2)}_2 \ , A^{(3)}_4 = A^{(1)}_2 A^{(2)}_2 \  .
\eeqs
\beq
Z_2 =\hat N'_1 + m'^2_1 - m^2_2 + (1-2x_1) M'^2 + (q^2+ q \cdot P) \frac{p'_{\perp}\cdot q_{\perp}}{q^2}.
\eeq


\end{appendix}


\newpage

\end{document}